%% file: conference101719.tex
\documentclass[journal,onecolumn]{IEEEtran}
\IEEEoverridecommandlockouts
\usepackage{cite}
\usepackage{listings}
\input{solidity-highlighting.tex}
\usepackage{amsmath,amssymb,amsfonts}
\usepackage{algorithmic}
\usepackage{graphicx}
\usepackage{textcomp}
\usepackage{xcolor}
\def\BibTeX{{\rm B\kern-.05em{\sc i\kern-.025em b}\kern-.08em
    T\kern-.1667em\lower.7ex\hbox{E}\kern-.125emX}}
\begin{document}

\title{A Survey on Vulnerabilities of Ethereum Smart Contracts}

\author{\IEEEauthorblockN{Zulfiqar Ali Khan and Akbar Siami Namin} \\
\IEEEauthorblockA{\textit{Department of Computer Science} \\
\textit{Texas Tech University}\\
\textit{zulfi.khan, akbar.namin@ttu.edu}}\\
}

\maketitle

\begin{abstract}
Smart contract (SC) is an extension of BlockChain technology. Ethereum BlockChain was the first to incorporate SC and thus started a new era of crypto-currencies and electronic transactions. Solidity helps to program the SCs. Still, soon after Solidity's emergence in 2014, Solidity-based SCs suffered many attacks that deprived the SC account holders of their precious funds. The main reason for these attacks was the presence of vulnerabilities in SC. This paper discusses SC vulnerabilities and classifies them according to the domain knowledge of the faulty operations. This classification is a source of reminding developers and software engineers that for SC's safety, each SC requires proper testing with effective tools to catch those classes' vulnerabilities. 
\end{abstract}


  \vspace{-0.2in}
\section{Introduction}
BlockChain is the most significant development to promote crypto-currencies. There are variations of BlockChain. Ethereum BlockChain, also known as Ethereum Virtual Machine (EVM) allows various unknown individuals to join hands and work together under a digital agreement known as a SC. Contracts require rules, but in this case, a programming language called {\it Solidity} embeds the rules within the SC itself. SC does not contain any "main" method, so it's not self-executable. 

SC is deterministic. This constraint generates the same output when any node of the Ethereum network executes the SC. Nodes can be users, or nodes can be miners responsible for validating SCs by solving a mathematical puzzle. The mathematical puzzle should generate the same result on all nodes, and this validation process follows the consensus protocol.

Once the mining process completes, its owner uploads the SC on the BlockChain. If the contract does not fit according to Ethereum rules, miners discard it. This process might follow the re-submission of SC. Thus the consensus protocol becomes a  method for developing trust among the parties. 

The trust is that there exists no error or fraud. But once the owner uploads the SC, it becomes immutable. Thus an unsafe SC can shatter the trust. Hackers can misuse it, causing considerable losses to the SC account holders. Therefore it is necessary to identify the vulnerabilities of SC before one uploads the SC on the BlockChain.

This survey research paper focuses on 20 vulnerability patterns. We have also provided the Solidity code corresponding to each vulnerability. We have used the context of SC's faulty operation for the classification of vulnerabilities. For brevity reasons, we have skipped the mitigation techniques. 

  \vspace{-0.05in}
\section{Motivation}
Previous research work has created several taxonomies related to vulnerabilities of SCs. The typical approach is to categorize the vulnerabilities based upon EVM, BlockChain, and Solidity associated issues as discussed in \cite{\iffalse 115, 26,\fi 114}. The above three are broader classes and give less information to the readers about the code's internal drawback.

Similarly, the survey in \cite{116} classifies the SC vulnerabilities in the context of EVM and Solidity. Other significant survey grouped the vulnerabilities based upon layering \cite{117} like application layer, data layer, and consensus layer. Furthermore, the work in \cite{118} used NIST Bugs Framework for the classification. However, SC introduced new kinds of vulnerabilities \cite{117} not common in traditional programming paradigms. 

We are the first to classify SC's security issues according to the vulnerable operation's domain knowledge. A good understanding of vulnerabilities requires coding examples along with description. Work in \cite{116, 117} also provided coding examples, but we address a different set of vulnerabilities. It is worth mentioning that the most relevant survey on the SC attacks with vulnerable SC code is available in \cite{9}, published in 2017. Since then, the Solidity programming language has undergone major changes. In this survey paper, we replicate vulnerabilities in Solidity code using the ``solc'' compiler version 5.0. The key contributions of this survey paper are:
\begin{itemize}
    \item[--] A classification of the SC vulnerabilities according to the domain knowledge.
    \item[--] A description of SC vulnerabilities through updated sample code.
    \item[--] A discussion of twenty SC vulnerabilities along with twenty-four SC vulnerability detection tools. 
    \item[--] A discussion on the regeneration of deprecated {\tt Var} vulnerability (section \ref{sss:Iof_coog}).
\end{itemize}

The remainder of this paper is organized as follows: section \ref{sec:Classification} presents our classification of SC's vulnerabilities. Section \ref{sec:Description} presents the description of SC  vulnerabilities with respect to domain knowledge along with some sample code. Section \ref{sec:smartcontracttools} lists the testing tools developed for SCs. Section \ref{sec:conclusion} concludes the paper and highlights the future research directions.

\section{A Classification of SC's Vulnerabilities Based on Domain Knowledge}
\label{sec:Classification}

SCs are susceptible to coding threats. One research classifies the vulnerabilities into security, functional, operational, and developmental categories \cite{95}.  These categories do not necessarily reflect any knowledge specific to SC. 

Our research classifies the vulnerabilities using the domain knowledge of operations performed by vulnerable SC. For instance, SCs communicate with each other by invoking methods. This inter-SC communication operation can pave the way for several coding irregularities like reentrancy, denial of service, mishandled exception, and so on. In the same way, other SC specific operational knowledge helped us to create vulnerability categories like Contractual, Arithmetic, Gas Related, Transactional, Randomization and Deprecated vulnerabilities. The domain knowledge-based classification provides (i) more knowledge about the cause of vulnerability (as compared to previous taxonomies discussed above) and (ii) we can drill down to the specific instructions (in some cases). For instance, in case of inter-contractual vulnerabilities, we can infer that {\tt call} is behind this vulnerability. Due to space limits, we have excluded randomization and deprecated vulnerabilities from the discussion.  

\begin{figure}[!t]
  \centering
  \includegraphics[width=\linewidth]{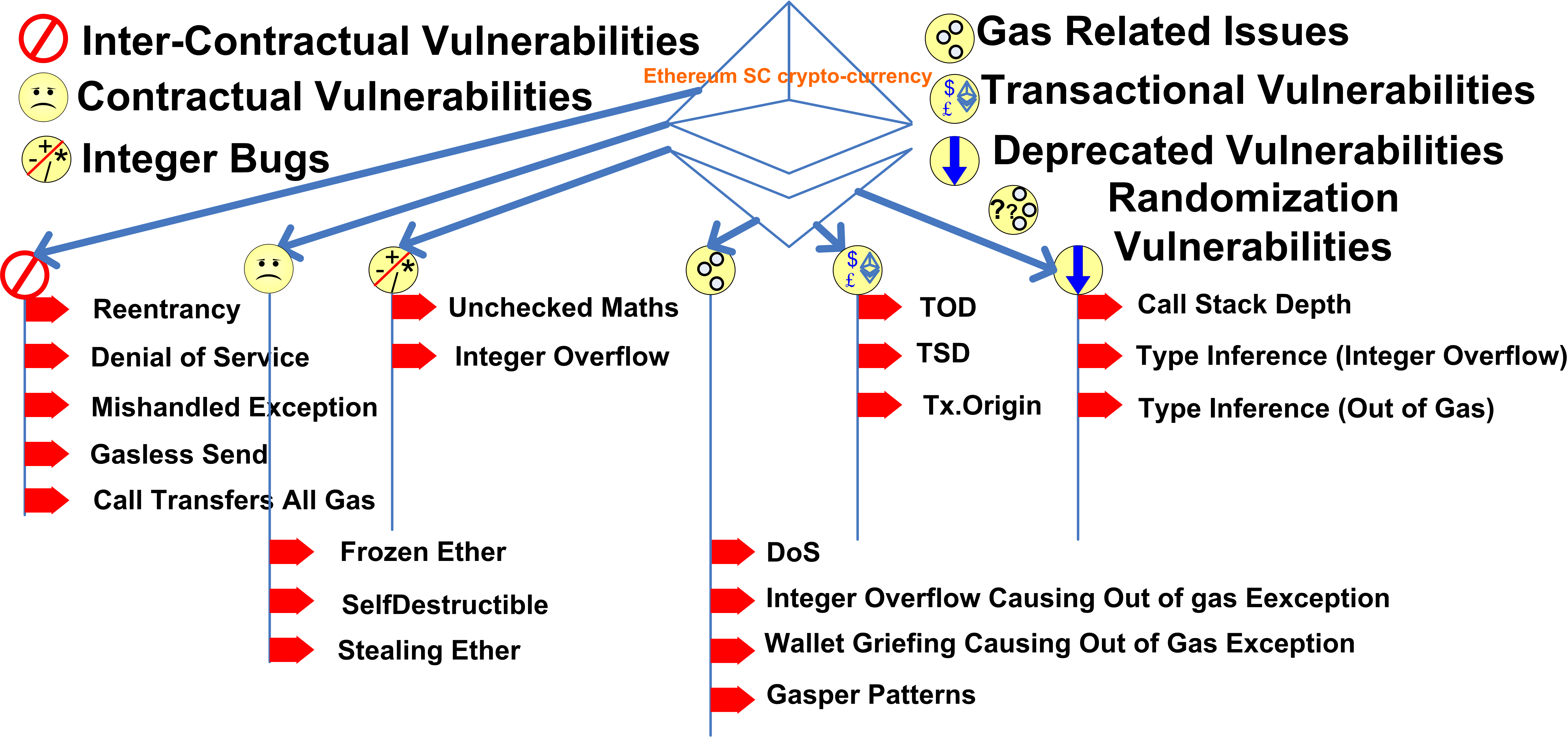}
  \caption{Classification of Ethereum SC vulnerabilities into seven categories.}
  \label{fig:CV}
    \vspace{-0.2in}
\end{figure}

Figure \ref{fig:CV} depicts our classifications of SC's vulnerabilities. As shown in Figure \ref{fig:CV}, the SC's vulnerabilities are grouped into 1) inter-contractual, 2) contractual, 3) integer bugs, 4) gas related, 5) transnational, 6) deprecated, and 7) randomization vulnerabilities.  

\section{Descriptions of SC Vulnerabilities}
\label{sec:Description}

\subsection{Inter-Contractual Vulnerabilities}
Listing \ref{lst:reentrancy}\footnote{This paper lists the EVM opcodes in capital letters and references the Solidity functions and contents of Listings in this {\tt font}.}  shows an example of inter-contractual communication  between {\tt ModifiedBank} and {\tt ModifiedMalicious} SCs using {\tt call} modified from \cite{15} in the context of a bank and an attacker. 
{\tt call} can also generate reentrancy attack. 
Solidity provides {\tt send}, {\tt transfer}, and {\tt call} functions for contract-wide Ether transfers along with an external {\tt payable} fallback function (FF) at the receiving SC. FF is an anonymous function, which retrieves the transferred Ether in the global state variable {\tt msg.value}. Transferring Ether incurs gas charges. {\tt send} and {\tt transfer} can compensate for 2300 \cite{114} amount of gas which is sufficient only for executing FF not linked to state changes. {\tt call} can transfer entire gas, which can initiate reentrancy attack, as discussed below:

\begin{lstlisting}[frame = single, numbers=left, breaklines, tabsize=2, language=Solidity, label={lst:reentrancy}, caption={Example Solidity Code of Reentrancy, modified from \cite{15}.}]
pragma solidity ^0.5.1;
contract ModifiedMalicous{
	   ModifiedBank mb;
	   constructor(address payable addressOfBank, uint amount) public{
	   mb=ModifiedBank(addressOfBank); 
	   mb.withdraw(amount);
	   }
	   function() external payable{ mb.withdraw(msg.value);}
}
contract ModifiedBank{
	 mapping(address=>uint)bal;
	 //more
	 function withdraw(uint _amount) public returns(bytes memory message) {
	 if(bal[msg.sender]>= _amount){
	 (bool success, bytes memory returnsMessage) = msg.sender.call.value(_amount)());
	 bal[msg.sender]-=_amount;
	 /*more*/}}
	 function() external payable{}}
\end{lstlisting}

\subsubsection{Reentrancy}\label{subsubsec:reentrancy} Reentrancy is a devastating attempt to deprive the investors of their precious cash. Even though the reentrancy attack (or the DAO attack \cite{94}) did not knock out the newly born SC technology, but created a stir in the Ethereum community. Reentrancy problem soon became an exciting topic of discourse among the BlockChain researchers because of two reasons: its inherent characteristics (discussed below) and the enormity of Ether flown out from SC as a result of it. 

Reentrancy problem occurs when an attacker reenters the SC repeatedly. This reentrance sparks of multiple issues like fueling out the entire gas of the victim, hijacking of the victim's SC , execution of two functions at the same time and undue transfer of funds from victim’s SC to the attacker’s account. 
Developmental issues coupled with the built-in implicit invoking nature of FF provide a stimulus for this attack. 

\begin{figure}[!t]
  \centering
  \includegraphics[width=\linewidth]{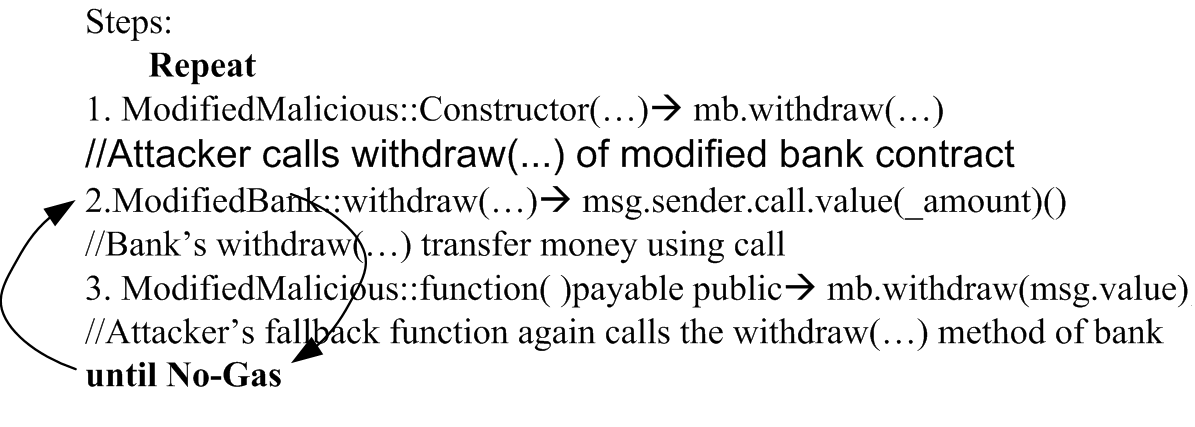}
  \caption{The reentrancy cycle, (Solidity-like) code modified from \cite{15, 47}.}
  \label{fig:inter-contractual}
      \vspace{-0.2in}
\end{figure}

There are several variants of this type of attacks like same or cross-function reentrancy (\cite{50}), ``Single-entrancy'' (\cite{39}), and the state-of-the-art reentrancy attacks discussed in \cite{83}. Reentrancy is a repeated operation and the statements execute in a cycle as shown in Figure \ref{fig:inter-contractual}. As discussed above, Listing \ref{lst:reentrancy} tries to simulate the banking application. However, the banking contract provides only the {\tt withdraw(...)} method with a flaw. The attacker {\tt ModifiedMalicious} exploits this flaw to drain the Ether from the bank into its own account.  

Listing \ref{lst:reentrancy} demonstrates how the {\tt ModifiedMalicious} contract exploits the flaw in line\#16. Firstly, the attacker’s SC {\tt ModifiedMalicious} calls the {\tt ModifiedBank} SC’s {\tt withdraw} method in line\#6 to retrieve balance {\tt amount} from her SC account. As the SC {\tt ModifiedBank} executes line\#15 (using {\tt call}) for transferring, the action results in the voluntary invoking of the costly FF of the attacker in line\#8. It is costly because the FF in line\#8 is not empty and contains the code to invoke the withdraw method of {\tt ModifiedBank} in line\#8. This process repeats until the bank or the attacker reaches out of gas state. However, in the meantime attacker may drain handsome amount from the bank because the debit process never deducts her balance, i.e., line\#16 never executes. 

\subsubsection{Denial of Service (Unexpected {\tt throw)}}\label{sss:DoS_ut} Denial of Service occurs due to various reasons. External functions may contain broken linkage due to the use of {\tt throw} (deprecated). Listing \ref{lst:DoS} (excerpted from \cite{9}) shows the contracts (i.e., {\tt MKotET1}, {\tt MKotET1\_1}). Both  are related to a game to acquire the {\it throne of KingofEther}. The throne's price is the money required by current king to leave the throne along with some processing fee to the SC's owner. However both the contracts suffer from Denial of Service vulnerability. {\tt MKotET1} (starting from line\#1)  exhibits Denial of Service threat using {\tt transfer} in line\#7 and {\tt MKotET1\_1} (starting from line\#12) displays Denial of Service threat using {\tt call} in line\#17.  {\tt MKotET1} and {\tt MKotET1\_1} use {\tt transfer} and {\tt call} respectively to deliver Ether to SC {\tt MMallory} (starting from line\#22), which contains a risky FF, in line\#23, which reverts in line\#24, throwing an exception unconditionally. Both {\tt transfer} and {\tt call} fail to deliver Ether to ‘{\tt MMallory’} due to the {\tt revert} in line\#24.


\begin{lstlisting}[language=Solidity, label={lst:DoS}, caption={KingOfEtherThrone threat variants:Transfer, Call \cite{9}.}]
	contract MKotET1{
	   address payable emperor; uint public rewardPrice = 500;
	     //declarations for MfindCrownPrice( ) & MNRewardPrice( )
	   function( ) external payable {
	     require (msg.value >= rewardPrice);
	     uint MCrownPrice = MfindCrownPrice();
	     emperor.transfer(MCrownPrice);
	emperoror = msg.sender;//Unreachable
	     rewardPrice= NRewardPrice();//Unreachable
	   /*more*/}//MKotET1 ends
	//modified KotET using Call
	 contract MKotET1_1{
	// some declarations
	    function( ) external payable{
	      require(msg.value  >= rewardPrice);
	      uint MCrownPrice = MfindCrownPrice( );
	      (bool success,) = emperor.call.value(MCrownPrice)("");
	      require(success);//throw if !success
	      emperor= msg.sender;//Unreachable
	      rewardPrice = NewRewardPrice();//Unreachable
	/*more */}//MKotET1_1 ends
	contract MMallory {   
	function ( ) external payable { 
	revert( ); } }//MMallory ends
\end{lstlisting}

\subsubsection{Mishandled Exception}\label{subsubsec:mshe_ucs} Mishandled Exception \cite{11} is a frequently occurring threat \cite{16} and appears by other names in the surveyed literature such as ``{\it Unchecked send}'', ``{\it Unchecked External Call}'', and ``{\it Exception Disorders}.'' Exceptions are run-time errors. One of the well-known exceptions in the Ethereum network is the out-of-gas exception. However, when a SC invokes an untrusted external function, a programmer must take extra care. For instance, sending Ether, implicitly invokes the FF of another SC. FF may fail and the reason might not be the out-of-gas exception \cite{36}. 

    In Solidity, we can use  {\tt transfer}, {\tt send}, and {\tt call} for sending Ether to another SC. If an exception occurs in the callee, {\tt transfer} propagates the exception in the caller's SC, which is safe. APIs like {\tt call}, {\tt send}, and {\tt delegatecall} return false and the execution continues \cite{7 }. Thus, if the programmer skips the checking of the false returned value, execution would continue resulting in an inconsistent state \cite{79}. Surveyed literature argues that the owner deliberately does not {\tt throw} an exception if the {\tt send} operation fails. This attitude may result in exceeding the call-depth stack (i.e. CDS, which is a deprecated vulnerability) \cite{9, 61}. {\tt MMallory} contract in Listing \ref{lst:exception} undo the transfer of Ether to itself. The {\tt payable} directive facilitates the importing of Ether, as discussed in section \ref{subsec:contvul} under frozen Ether vulnerability. However, anybody sending Ether to {\tt MMallory} will suffer from exception due to {\tt revert} in line\#2.

\begin{lstlisting}[language=Solidity, label={lst:exception}, caption ={'revert' in line\#2, causes mishandled exception \cite{9} to the caller.}]
	contract MMallory{	   
	function() external payable { revert (); }}
\end{lstlisting} 

\subsubsection{Gasless {\tt send} \cite{11}} \label{subsubsec:gaslessSend}{\tt send} is associated with a fixed gas stipend of 2300 \cite{47}, which is enough to execute an empty FF. If the FF modifies the state of the SC, then the required gas can increase beyond 2300. This costly FF results in an out-of-gas exception \cite{9}. Costly FF can be due to the developer’s mistake instead of a malicious activity \cite{21}. In other words, if the ``gas used'' is high, an exception occurs, and the malicious miner may keep the untransferred amount.

\begin{lstlisting}[language=Solidity, label={lst:gasless}, caption={\iffalse Receiver Contract's FF changes state variable
'TotalBal', line\#7, which makes it a costly FF \fi
Receiver  SC’s  FF is costly: alters ’TotalBal’, line\#7, \cite{107}.}] 
  contract Sender { 
    function pay(uint val, address payable _recv)public { //_recv points to contract Receiver
        if(_recv.send(val)){ } else { }  }}
  contract Receiver {
    uint public TotalBal = 0;//state variable
    function() external payable { 
        TotalBal = TotalBal + msg.value; } }
\end{lstlisting}

But in some cases, the malicious SC owner may keep the non-transferred amount and miner only gets his fee. {\tt call} mitigates this threat but transfers all gas which causes reentrancy. Listing \ref{lst:gasless} shows two SCs, {\tt Sender} line\#1 and {\tt Receiver} line\#4 modified from \cite{9}. {\tt Sender} transfers funds to Receiver using {\tt send} in line\#3 but {\tt Receiver} retrieves the Ether using a costly FF in line\#6. Transfer increments the state variable {\tt TotalBal} in line\#7. But transfer in line\#3 becomes a threat as {\tt send} uses 2300 amount of gas, which is not enough to execute a costly FF.

\subsubsection{{\tt call} Transfers All Gas} At the bytecode level, Solidity’s  {\tt send} and  {\tt transfer} translate into CALL (EVM bytecode). CALL executes the FF of the calling SC and can fail due to insufficient gas. It consists of four stack arguments: 1) the amount of gas required for the transaction, recipient’s address, 2) the amount of Ether to be transferred 
, 3) size, and 4) the location of input data along with the size and location of return data \cite{37, 55}. {\tt call} can provide the stimulus for reentrancy attack by forwarding all the gas \cite{95}. However, implicit {\tt call} statements are also possible. Listing \ref{lst:implicit_call} shows the code modified from \cite{111}. Line \#7 uses {\tt call} implicitly, which is vulnerable; the first parameter is the Ether value, while the second parameter is the argument of {\tt f(..)}.

\begin{lstlisting}[language=Solidity, label={lst:implicit_call}, caption={\iffalse Contract test2 uses implicit CALL to send amount (Ether) to function {\tt f()}, in line\#7\cite{111}\fi {\tt test2}: implicit CALL sends amount to {\tt f()}, in line\#7 \cite{111}.}]
 	contract test1 {
	    function f(int x) public payable returns (int){ return x;} }
	contract test2 {
	    function testA(address _add1) public { 
	        test1 a = test1 (_add1);
	        uint amount = 500;
	        a.f.value(amount)(2);//vulnerable
	        a.f.value(amount);//non-vulnerable
	        } }
\end{lstlisting}

\subsection{Contractual Vulnerabilities}
\label{subsec:contvul}

Contractual bugs impact the SC itself. Both the attacker and owner exploit them for causing harm to the SC users. The owner can design the SC to prevent leakage (or transfer) of any Ether, thus turning the SC into a black hole. However, the worst happens when an attacker uses the unprotected {\tt selfdestruct} command to destroy the SC. The attacker enjoys the balance of the account if the attacker changes the ownership of SC before destroying it.

\subsubsection{Frozen Ether}\label{subsubsec:frozenEther} Also known as ``{\it Locked Money}'' \cite{95} or ``{\it be no black hole}'' \cite{17}. The frozen Ether threat deprives the SC account holders of Ether worth millions of dollars, as in the case of parity wallet SC. One atypical impact of the above exploits resulted after the accidental killing of the library SC, which provided an external route to Ether in parity Wallet SC (and to other {\tt multisigWallet}-like SCs \cite{76}). 

Apart from the accidental execution of suicide command, which the hobbyist confessed of doing in his issue\# 6995 on Github\cite{113}, there could be coding loopholes in the SC, which can prevent exporting of Ether from the SC. Programmatically, exporting Ether weakness applies to SCs, which lack statements like {\tt call}, {\tt send}, or {\tt transfer}, which move the funds outside the SC, along with the presence of {\tt payable} directive in the SC, which on the other hand facilitates importing of Ether. To summarize, ``frozen Ether'' vulnerability occurs when:

\begin{enumerate}
\item SC permits inbound Ether traffic but shuts the outbound Ether traffic. Reference \cite{76} labels such SCs as greedy. FF handles inbound Ether traffic but {\tt call}, {\tt send} and {\tt transfer} handle outbound Ether traffic. Listing \ref{lst:Payable} uses a SC modified from \cite{17}, which shows the freezing Ether vulnerability because the SC contains a method having {\tt payable} directive in line\#2 but does not contain program paths leading to CALL, DELEGATECALL, or SELFDESTRUCT opcodes:

\begin{lstlisting}[language=Solidity, label={lst:Payable}, caption={{\tt payable} directive creates an Ether receiving FF in line\#2.}]
	contract ModifiedBitway{
	   function ( ) external payable { } }
\end{lstlisting}

\item Wallet SC relies on another SC or library. The library SC provides functions to support the Wallet, for instance, the library can provide function for transferring Ether. But if the library SC eventually kills herself by executing {\tt selfdestruct} command or some other SC (or an attacker) accidentally (or deliberately) kills the library SC, then it would close the doors of Ether extraction from the wallet SC. Listing \ref{lst:Withdraw}, modified from \cite{100}, shows the use of {\tt delegatecall} in line\#4 to load the code from the address {\tt 0xNewLibrary}, line\#2, containing the library's {\tt withdraw()} method.

\begin{lstlisting}[language=Solidity, label={lst:Withdraw}, caption={withdrawM() uses library through hardcoded address \\(line\#2) and {\tt delegatcall} (line\#4) 
using that address \cite{100}.}] 
	contract testDC{
	      address _nl = 0xNewLibrary;
	      function withdrawM() {   
	         _nl.delegatecall(msg.data); } }
\end{lstlisting}

\end{enumerate}

\subsubsection{Self-Destructible}\label{sss:sd} {\tt selfdestruct} (previously known as {\tt suicide}) allows a SC to destroy itself by releasing the Ether of account holders. SC uses this alternative in emergencies. Research conducted by \cite{76} terms a SC suicidal (i.e. vulnerable for SELDESTRUCT opcode) if the SC does not correctly guard the {\tt selfdestruct} command. Attacker requires two things to kill a SC: (1) reaching into the conditional statement enclosing the {\tt selfdestruct} statement, and (2) attaining the ownership of SC.

{\tt selfdestruct} causes unresponsiveness \cite{94} of the SC resulting in ``Denial of Service.'' {\tt selfdestruct} deletes the SC's code permanently \cite{18, 89}. All the SC's funds would transfer to the account associated with SC. This transfer will not trigger FF \cite{18}. A beneficiary can be an existing account or the account may not exist. In the latter case, the destruction process creates the account and charges fees for it \cite{37}. 
Contract {\tt MDiscontinue} in Listing \ref{lst:Selfdestruct} provides a {\tt TerminateMe} function to destroy the contract with unprotected {\tt selfdestruct} (i.e., without any {\tt if} block) in line\#6, modified from \cite{75}: 

\begin{lstlisting}[language=Solidity, label={lst:Selfdestruct}, caption={\iffalse {\tt selfdestruct}, line\#6, without any guard makes SC a suicidal\cite{75}\fi{\tt selfdestruct}, line\#6, without any guard, suicidal SC \cite{75}.}] 
contract MDiscontinue{
  address payable owner;
  constructor() public {   
    owner = msg.sender; } 
    function TerminateMe() public {
      selfdestruct(owner);}}//suicidal
\end{lstlisting}

\subsubsection{Stealing Ether}\label{sss:se} Stealing Ether vulnerability (or Unsecured Balance) surfaces when the SC initializes the owner field indirectly as in line\#4 of Listing \ref{lst:stealing} in a function other than the constructor. The use of a function for initialization of owner address (as in line\#3) can lead to a problematic situation. This situation is similar to parity SC, which doomed the multi-signature SC of 30m dollars \cite{4}.

\begin{lstlisting}[language=Solidity, label={lst:stealing}, caption={\iffalse State variable 'owner' initialized in line\#4, outside constructor\cite{100}\fi{\tt owner} initialized in line\#4, outside constructor\cite{100}.}]
contract CompWallet{
  address payable owner;//state variable
  function initComWallet(address payable _owner) public{  
    owner = _owner;}//any user can change owner
    function withdraw(uint _amount) public {
      if (msg.sender == owner){
        if(!owner.send(_amount)){}
          else {}}}}
\end{lstlisting}

\subsection{Arithmetic Bugs}
\label{intbug} 

This problem occurs as a result of a mathematical operation. Most significant bug is the integer overflow/underflow, which is a common problem in programming languages. Increments beyond the maximum or decrements below the lowest value (i.e. wrap-around) may generate wrong results. Thus, developers must perform manual checking (i.e., employ a human expert to check the code); otherwise, code may create a wrap-around error. One solution is to use {\tt SafeMath} library. The latest research recommends using the ``{\it solc-verifier}'' tool. 

\subsubsection{Integer Overflow/Underflow}\label{subsubsec:intof_uf} 
\begin{figure}[h]
  \centering
  \includegraphics[width=\linewidth]{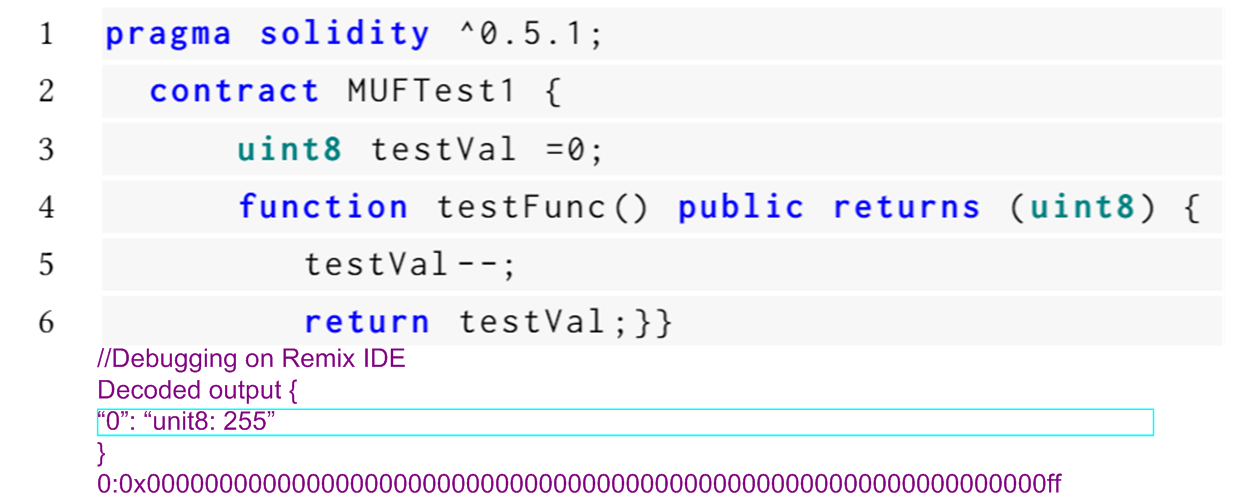}
  \caption{Underflow SC, MUFTest1, modified from \cite{108}, and the result of debugging it on REMIX IDE.}
\label{fig:overflow_underflow}
\end{figure}

Ethereum has nothing to do with data types leaving the compiler responsible for catching integer overflows and under-flows \cite{79, 99}. Solidity is rich in integer data types with flexible sizes (uint8, uint16, uint24,..uint256, int8, int16, int24,..int256) but does not support floating-point math \cite{95}. Thus, the diversity of integer data types provides no benefit to SC. 
Work in \cite{1, 79} provides an example code generating integer bugs. Figure \ref{fig:overflow_underflow} shows an underflow SC and its output on Remix (0 changes to 255 on decrements). 

{\tt SafeMath} \cite{33, 95} provides several functions to replace the ordinary arithmetic operations in SCs. Listing \ref{lst:safemath}, lines\#2-4, show the logic of {\tt SafeMath} library's  method {\tt sub(..)}. Figure \ref{fig:overflow_underflow} SC, {\tt MUFTest1}, is modified from \cite{108}. We replaced the line\#5 in Figure \ref{fig:overflow_underflow} by {\tt sub(..)} method of SafeMath library as shown in Listing \ref{lst:safemath}, line\#9, of SC {\tt MUFTest2}.

\subsubsection{Unchecked Maths} Unchecked math means that a SC is not using strategies to protect mathematical statements from overflows/underflows. A good practice is to protect the code using assertions and {\tt SafeMath} library as shown again in Listing \ref{lst:safemath}, line\#9.

\begin{lstlisting}[language=Solidity, label={lst:safemath}, caption={Use of {\tt sub(..)} function, line\#9, to avoid underflow\cite{108}.}]
	library SafeMath {
	   function sub(uint8 x, uint8 y) internal pure returns (uint8) {
	      assert(y <= x);
	      return x - y; } }
	contract MUFTest2 {
	using SafeMath for uint8;
	   uint8 testVal= 0;
	   function Utest() public returns (uint8){
	      testVal= testVal.sub(1);
//instead of : testVal- - (in line#5, Fig. 3)  
	      return testVal; } }
\end{lstlisting}

\subsection{Gas Related Issues}
\label{subsec:outofgasexc}

The gas serves two essential purposes for the EVM network. Firstly, gas serves as compensation to the miners’ efforts for recording transactions. Secondly, the gas acts as a fuel for running a transaction and thus prevents long transactions from hijacking the EVM scheduling scheme. Logically, it means that if the user does not pay enough gas fee as required for the transaction, the transaction will fail by generating an ``{\it out of gas}'' exception. Surprisingly, an integer overflow can also cause ``out of gas.'' Other examples are Denial of Service, and Wallet Griefing, as discussed below:

\subsubsection{Denial of Service (Costly Loops Causing Out of Gas Exception)}\label{sss:DoS_clc_oog} EVM protects programs from Denial of Service attacks by forced termination. EVM allocates gas at the start of execution, and each execution step results in some deduction. If the remaining amount after deduction is less than the amount required for execution, EVM \cite{36} terminates the SC's execution, causing partial or full rollback \cite{44}. The termination can occur even without the influence of an attacker \cite{71}. 

An attacker can manipulate the {\tt arr}, line\#4, in Listing \ref{lst:gascost} by adding additional addresses. Thus, increasing the execution cost and transfer to manipulated addresses. In the worst case, the entire gas may be exhausted, resulting in full {\tt revert}. Denial of Service can occur due to the presence of {\tt revert} in external function, as in Listing \ref{lst:DoS}, line\#24.

\begin{lstlisting}[language=Solidity, label={lst:gascost}, caption={\iffalse A gas costly For loop traversing an unsized array, creates an unbounded mass operation\cite{36}\fi Unbounded mass operation: traversing an unsized array\cite{36}.}]
	contract testLL{
	   uint constant LARGEGAS = 100000; 
	   address payable addrArr;
	   function LongList(uint256 memory nextV, uint[ ] memory arr, address payable _addr) public {
	      uint256 i= nextV;
	      addrArr = _addr;
	      for ( ;i < arr.length && gasleft() > LARGEGAS; i++) {
	         addrArr.send(arr[i]); }
	      nextV = i; } }
\end{lstlisting}

Another example is the Denial of Service due to a costly {\tt for} or a {\tt while} loop. This may cause depletion of gas in each iteration and finally resulting in Denial of Service. Work presented by \cite{36} renames this vulnerability, as ``{\it Unbounded mass operations} due to an unsized array variable {\tt arr} in line\#7, Listing \ref{lst:gascost}, and \cite{36} have proposed resumable loops. Function {\tt LongList(..)}, line\#4, in Listing \ref{lst:gascost}, is excerpted from \cite{36}. In the case of {\tt revert}, due to ``out of gas'' threat, {\tt nextV}, in line\#9, points to the {\tt arr} index, from where to resume.

\subsubsection{Integer Overflow (Causing Out of Gas Exception)}\label{sss:Iof_coog} Authors of SmartCheck \cite{95} and MadMax \cite{36} discuss about the integer overflow issue in a loop in the context of {\tt Var} (deprecated, Solidity used Var for Type Inference \cite{50, 95}). However, an integer overflow can occur even when {\tt Var} does not determine the type of loop index (i.e., {\tt uint8}) variable. By this, we mean that integer overflow can occur if the type of loop index variable is a short integer at run-time. Listing \ref{lst:integer} shows an overflow without using {\tt Var}, as in line\#4 in Listing \ref{lst:integer} when the loop counter variable becomes greater than or equal to 255. 

\begin{lstlisting}[language=Solidity, label={lst:integer}, caption={Short integer overflows even without {\tt Var }\cite{109} in line\#4}.]
contract Overflow {
  int [300] emp;
  function testOF() public returns (bool) {
    for (uint8 i = 0; i < emp.length; ++i) { }}}
\end{lstlisting}

\subsubsection{Wallet Griefing Causing Out of Gas Exception}

This vulnerability occurs if a SC uses a loop to send Ether to multiple SCs. Thus, if one receiver fails, then the entire transaction fails. The receiver can fail due to a bad FF (line\#2, Listing \ref{lst:exception}). If the sending SC uses a {\tt throw} (i.e., {\tt revert}) to handle the failure of {\tt send} then it can exacerbate the situation because {\tt throw} consumes the entire gas \cite{16, 43}and locks the sender's SC \cite{28, 36}. Repeated attempts may also fail due to ``out of gas'' situation, as in line\#6, Listing \ref{lst:Griefing}, modified from \cite{36}. However, the latest version in can achieve the same effect by using {\tt require} and {\tt transfer} instead of {\tt send} and {\tt revert}.

\begin{lstlisting}[language=Solidity, label={lst:Griefing}, caption={Wallet Griefing using {\tt revert}, {\tt send} fails in line\#6, \cite{36}.}]
for ( uint i = 0; i < employee.length ; i++) {
  if ( employee [i].paid < min_salary ) {
  // sentinel for making a payment.
  // code may lock the SC
  // due to {\tt revert} consuming all gas.
  if (!( employee [i].addr.send ( employee [i].bonusAmount ))) revert() ;
    employee [i] = newEmployee ; } }
\end{lstlisting}

\subsection{Transactional Irregularities}

EVM transactions become a source of greed for the miner, which validates them. Thus some miners can influence the transactions resulting in vulnerabilities like Transaction Ordering Dependence and Time Stamp Dependence.

\subsubsection{Transaction Ordering Dependence (TOD)}\label{sss:TOD} TOD is also known as ``{\it front running race condition}'' \cite{7}. This attack, ``selfish mining attack'', occurs due to the mishandling of transaction queue by miners. The owner/user incentivizes the miner to change the order of the transaction \cite{61}. 

One example of TOD is the case of marketplace SC as shown in Listing \ref{lst:ordering} modified from \cite{61} in which a miner may not honor a leading buyer's request at the cost of some other transaction. The buyer sends the transaction to buy at cost 100, line\#7, Listing \ref{lst:ordering}. At the same time, owner sends the transaction with a high gas fee to increase the price, line\#4, Listing \ref{lst:ordering}. The owner's transaction executes first due to the higher gas price incentive. The buyer's transaction completes next but the buyer pays more.

\begin{lstlisting}[language=Solidity, label={lst:ordering}, caption={Miner alters lines\#5-8, causing TOD: marketplace\cite{61}.}] 
	contract MMarketPlace {
	   uint private cost=100 ;
	     uint private inventory= 100; //more declarations
	     function incPrice ( uint _incCost ){
	        require ( msg.sender == owner )
	       cost = cost + incCost ; }
	   function buy ( ) returns ( uint ){
	     require(msg.value == cost);
	     require(inventory > 0));
	     inventory -= 1; /* more */ } }//use of SafeMath recommended
\end{lstlisting}

Contrary to the general notion about a miner in connection with TOD, the recent research conducted by \cite{79} argues that it is hard to exploit TOD threat because the attacker should be a miner, and there are less financial gains. Apart from this, there are some other programming problems. State variables often have a dependency on the function, which changes their values. Thus, coding such a function when multiple users are invoking that function is a concurrent programming issue. 

The concurrency in SCs requires some mechanism like semaphore to control access to the state variable. Reference \cite{50} argues that Solidity does not support concurrency. Thus, manipulation by a miner is a limitation of BlockChain rather than a bug. However, EVM must provide some solution for miner's problems, as the miners also contribute to immense power consumption. One solution proposed by \cite{34} is to delegate miners' role to a SC. Another name for this problem is the unpredictable state problem. This is because multiple invocations of a dependent function make it difficult to predict what the state and the values stored within a SC will be when a user executes the function.

\subsubsection{TimeStamp Dependence \cite{11}}
\label{subsubsec:tsd}Each block within the BlockChain contains three pieces of information: 1) timestamp, 2) cryptographic hash, and 3) the transaction data. 

\begin{itemize}

\item [--] Timestamp represents the time when the miner verifies all the transactions within the block after computing the proof-of-work puzzle. A miner can manipulate the block timestamp but has to complete the validation within 900s \cite{61}; otherwise other miners would reject the block. 

\item [--] Cryptographic hashesare deterministic functions. Miners exploit the deterministic hash values to verify the integrity of the block's data. Cryptographic hash chains the current block with the previous block as there is a dependency between hash values of the two said blocks. 

\item [--] Transaction data can vary based upon transactions. Still, for the most straightforward transactions between two SC accounts, the transaction data would be the sender's and receiver's SC account addresses and the amount of Ether sender transfers to the receiver.
\end{itemize}

Despite the doubtful accuracy, SCs use the timestamp for random number generation. Due to the miner's involvement in setting the timestamp's value, the timestamp becomes a so-called deterministic random value. Hence, the usage of the timestamp as a random value in lottery implementation is vulnerable \cite{37}. Programmatically, {\tt block.timeStamp} retrieves the timestamp associated with a block. However, one can just use {\tt now} (alias for block.timestamp) also to retrieve the timestamp, as in Listing \ref{lst:timestamp}, lines\#3-4. Solidity uses {\tt now} for simplicity and bytecode implementation. But Solidity does not discriminate between {\tt now} and {\tt block.timeStamp} and hence both are vulnerable.

\begin{lstlisting}[language=Solidity, label={lst:timestamp}, caption={Miner's misuse of {\tt now} (lines\#3-4) causes TSD.}] 
	contract TSD{
	   function Mpay () public{
	      uint tTime = now; 
	         if (tTime > (now + 2) ){ if(!msg.sender.send(200)) { } else { } } } }
\end{lstlisting}

\subsubsection{tx.origin} 
{\tt tx.origin} is a transaction state variable which indicates the originator of the transaction. Other transaction state variables also exist like {\tt tx.GasPrice}. But {\tt tx.GasPrice} has a fixed value so the adversary cannot change it \cite{50}. On the other hand, {\tt tx.origin} can vary. This variation can lead to attacks because we cannot use {\tt tx.origin} to ratify the contract's owner, as shown in Listing \ref{lst:origin}.

Lines\#1-9 in Listing \ref{lst:origin} show the victim's SC (i.e., {\tt TxUserWallet}) and lines\#10-11 show the interface (i.e., {\tt TxUserWallet}). Note that the names of the victim's SC and the interface are the same but they are in different files. 

\begin{lstlisting}[language=Solidity, label={lst:origin}, caption={Example  of  tx.origin\cite{77},  Victim’s  SC  (i.e. {\tt TxUserWallet}).}]
contract TxUserWallet {
  address owner;
  constructor() public  {
    owner = msg.sender; }
  function sendTo(address dest, uint256 amount) payable public returns (bytes memory theMessage)
    {require(tx.origin == owner);
	(bool success, bytes memory returnMessage) = dest.call.value(amount)();
	require(success);
	return returnMessage; } /* more */}
interface TxUserWallet {
  function sendTo(address dest, uint amount) external;}
contract TxAttackWallet {
  address owner;
  constructor()  public  {
    owner = msg.sender; }
    function() external payable  {
      TxUserWallet(msg.sender).sendTo(owner, msg.sender.balance); } /* more */}
\end{lstlisting}
The rest of the code from lines\#12-17 show the attacker's SC, {\tt TxAttackWallet}. The surveyed literature recommends the replacement of {\tt tx.origin} with {\tt msg.sender} particularly for authenticating the sender of a message \cite{16}. {\tt tx.origin} represents the address of the first account in the call chain (i.e., the list of calls related to the currently executing transaction), whereas {\tt msg.sender} is the original caller \cite{76, 95}.

\section{Tools for Testing SMART CONTRACTS}
\label{sec:smartcontracttools}

we provide a brief description of SC tools developed for detecting above mentioned vulnerabilities. We grouped the testing tools into dynamic and static-based tools.

\subsection{Testing Tools based on Static Analysis of SCs}
There are a good number of static-analysis for testing SCs:

\begin{itemize}
\item [--] {\it Zeus \cite{50}.} Zeus is a tool for formal verification of SCs using abstract interpretation and symbolic model checking. Zeus works directly on the high-level of SC code. Zeus detects threats like reentrancy (section \ref{subsubsec:reentrancy}), Unchecked and Failed {\tt send} (section \ref{subsubsec:mshe_ucs}), Integer Overflows (section \ref{subsubsec:intof_uf}), and Timestamp dependency(section \ref{subsubsec:tsd}).

\item [--] {\it VeriSolid \cite{69}.} VeriSolid is a SC development tool. VeriSolid uses a transition system model to generate Solidity-based formally verified SCs. Automatic code generation is an important achievement in the context of formal verification tools. VeriSolid prevents reentrancy (section \ref{subsubsec:reentrancy}) by design and uses liveness property to prevent Denial of Service (section \ref{sss:DoS_ut}, \ref{sss:DoS_clc_oog}).

\item [--] {\it Vandal \cite{16}.} Vandal is a static analysis tool. Vandal performs security analysis of EVM bytecode using a logic language, called Souffle, to transform the analysis into $C++$. Vandal's static analysis library functions detect threats like ``Unchecked Send'' (section \ref{subsubsec:mshe_ucs}), Reentrancy (section \ref{subsubsec:reentrancy}), UnSecured Balance (section \ref{sss:se}), Destroyable SC (section \ref{sss:sd})\cite{16}.

\item [--] {\it Teether \cite{55}.} 
Teether focuses on the same idea of critical paths as in \cite{17}. Teether constructs the CFG of the SC using the EVM bytecode, which helps to detect critical paths. The authors discussed the peculiar problem of backward traversal associated with JMP because of JMP's similarity with x86's return statement. 

\item [--] {\it SmartScopy \cite{31}.} SmartScopy is an attack synthesizer. SmartScopy performs summary-based symbolic evaluation, which reduces the program size for symbolic analysis (SA) required for automatic generation of adversarial SC. The adversarial SC confirms the presence of vulnerability in the victim SC, detected by manual analysis. SmartScopy detects threats like reentrancy (section \ref{subsubsec:reentrancy}), timestamp dependence (section \ref{subsubsec:tsd}), and Gasless send (section \ref{subsubsec:gaslessSend}). 

\item [--] {\it SmartCheck \cite{95}.} SmarkCheck helps to remove the simple bugs quickly. However, for removing non-trivial bugs, \cite{95} recommends using more sophisticated techniques like taint analysis (TA). The authors identified several coding threats like reentrancy (section \ref{subsubsec:reentrancy}, Mishandled Exception \ref{subsubsec:mshe_ucs}, {\tt call} transfers all gas, and so on. The SmartCheck \cite{95} related research provides a comprehensive list of threats based upon exploits related to security, functional, operational, and developmental issues.

\item [--] {\it Securify \cite{100}.} Securify is a static analysis tool, focusing on patterns (compliance or violation). The tool extracts the domain knowledge from patterns related to a security property. Securify detects threats like Stealing (section \ref{sss:se}) and frozen Ether (section \ref{subsubsec:frozenEther}), Reentrancy (section \ref{subsubsec:reentrancy}), Mishandled Exception (section \ref{subsubsec:mshe_ucs}) and Transaction Ordering Dependence (section \ref{sss:TOD}). Related violation properties with regarding threats are restricted write, Ether liquidity, no writes after calls, handled exception, restricted transfer and TOD \cite{100}.

\item [--] {\it Oyente \cite{61}.} Oyente performs static analysis of SCs. \cite{61} recommend the extension of Oyente and is a reality when one reads the details of the tools mentioned in \cite{2}. Oyente detects threats like TOD (section \ref{sss:TOD}), TimeStamp Dependence (section \ref{subsubsec:tsd}), and Mishandled Exception (section \ref{subsubsec:mshe_ucs}).

\item [--] {\it OSIRIS \cite{99}.} OSIRIS employs a strategy based upon taint analysis and symbolic execution and consists of an integer detection module. Symbolic execution module constructs a control flow graph (CFG) from the bytecode. The CFG processes different paths of the SC using symbolic values, as in Maian \cite{76}. Osiris \cite{99} detects arithmetic bugs (overflow/underflow (section \ref{subsubsec:intof_uf}) and division by 0), truncation bugs (converting from a larger to smaller data size, e.g., 64-bit data to 16-bit data), and signedness bugs (converting a signed integer to an unsigned integer of the same width and vice versa).

\item [--] {\it MadMax \cite{36}.} MadMax focuses on automatic detection of gas-focused threats. Like Vandal \cite{16}, MadMax performs the de-compilation of EVM bytecode and similarly uses a logic-based approach to produce a high-level representation of the program model. MadMax defines strategies for surviving out of gas conditions in the context of resumable loops (section \ref{sss:DoS_clc_oog}), loops bounded by induction variable, and dynamically bounded loops.

\item [--] {\it KFrameWork-EVM-Semantics (KEVM) \cite{44}}:KEVM is a semantic analysis tool based upon SE. KEVM's development framework integrates a semantic debugger and a program verifier. Tool encapsulates a gas analyzer that computes gas bounds during execution and can help in detecting ``Denial of Service''(section \ref{sss:DoS_ut}, section \ref{sss:DoS_clc_oog}) threats.

\item [--] {\it Interactive Theorem Provers (ITP) \cite{45}.} Interactive Theorem Prover combines the idea of theorem proving with testing. Initially, this tool presents the desired behavior of EVM in LEM \cite{73}. Authors used both the community-based test suits and interactive theorem provers like Isabelle/HOL to test their EVM definitions. The authors divided the formalization into deterministic and non-deterministic formalization. The basic assumption for non-deterministic formalization was to segregate the environment from the system. This helped to reason about the adversarial attack and to model the reentrancy attack (section \ref{subsubsec:reentrancy}). 

\item [--] {\it Gasper \cite{19}.} Gasper uses Oyente Engine to generate the CFG and identifies the code for optimization. \cite{19} points out seven gas costly-patterns. Due to shortage of space, we have not discussed Gasper patterns in this paper.

\item [--] {\it FSolidM \cite{67}.}FSolidM allows development of secure SCs using Finite State Machines (FSMs) \cite{7}. FSolidM uses a set of plugins and design patterns that developers can add to the SC for implementing locking, maintaining transaction counter, and enforcing timed transitions to safeguard against reentrancy (section \ref{subsubsec:reentrancy}), transaction ordering dependence (section \ref{sss:TOD}), and time constraint, respectively.

\item [--] {\it FVF*\cite{15}.} The research work in \cite{15} describes the verification of Ethereum SCs using F* (FVF* stands for formal verification using F*). 
The work in \cite{15} added an effect system in F* which helps in the detection of  ``unchecked send'' (section \ref{subsubsec:mshe_ucs}) and destructive patterns like reentrancy (section \ref{subsubsec:reentrancy}). 

\item [--] {\it EtherTrust \cite{38}.} Authors used the tool to prove the reachability property for SC's bytecode. EtherTrust ensures two things about the SC: (i) FFs should not result in DAO type of attack (section \ref{subsubsec:reentrancy}) (ii) data is not vulnerable to miner’s manipulation.

\item [--] {\it DappGaurd \cite{21}.} DappGaurd detects the diverse type of threats. However, instead of relying on bytecode or Solidity code, DappGaurd focuses on Transaction Receipts, which analyze live SCs, but the authors do not provide the source for retrieving TRs. DappGaurd's prototype version detects several threats, and for this purpose, DappGard incorporates the Oyente engine.

\item [--] {\it sCompile \cite{17}.} sCompile exploits the notion of ``critical paths'' i.e., in place of identifying a program to be vulnerable, sCompile identifies critical paths (i.e., money related inter-contractual paths involving {\tt call}) in the program. Developed in $C++$, sCompile uses Z3 SMT Solver for SE. sCompile detects threats like reentrancy (section \ref{subsubsec:reentrancy}), be no black hole (section \ref{subsubsec:frozenEther}, and unguarded {\tt selfdestruct} (section \ref{sss:sd}).

\end{itemize}

\subsection{Testing Tools based on Dynamic Analysis of SCs}

\begin{itemize}
\item [--] {\it Vultron \cite{103}.} Vultron is still in infancy stages and the authors have tested the prototype using truffle suite. Vultron is a test oracle, which stores bookkeeping information in variables. Vultron compares these variables with account balances related to SCs to determine the inconsistencies. Vultron can identify threats like reentrancy (section \ref{subsubsec:reentrancy}), exception disorder (section \ref{subsubsec:mshe_ucs}), integer overflow/underflow (section \ref{subsubsec:intof_uf}), and ``gasless send'' (section \ref{subsubsec:gaslessSend}).

\item [--] {\it Sereum \cite{83}.} Sereum modifies the ``goethereum'' client ``geth'' and adds an attack detector and taint engine. Sereum works at the bytecode level and the binary level does not keep type information. This fact makes it challenging to infer about the sensitivity of data\cite{75}. Sereum related research reports a reentrancy attack (section \ref{subsubsec:reentrancy}). 

\item [--] {\it Regaurd \cite{58}.} Regaurd is a dynamic analysis tool and incorporates a fuzzing based analyzer. Regaurd focuses on automatic detection of common threats in SCs like reentrancy bugs (section \ref{subsubsec:reentrancy}). Regaurd transforms the code into an intermediate representation (IR) (Abstract Syntax Tree). Finally, Regaurd executes the SC (with transactions as input) and forwards the dump of relevant operations of run-time analysis to the core detector to detect reentrancy bugs.

\item [--] {\it Maian \cite{76}.} Maian is a static analysis tool but also performs dynamic analysis of SC. For offline inspection, Maian's input is the EVM bytecode and the SC's initial state retrieved from the BlockChain. SA generates actual values for the transaction given SC's bytecode and analysis specification i.e., vulnerability category to search like Suicidal (same as {\tt selfdestruct},  \ref{sss:sd}) or Greedy (same as Frozen Ether, section \ref{subsubsec:frozenEther}) as input. Maian then uses the actual values in the validation step.

\item [--] {\it EasyFlow \cite{33}.} EasyFlow is a specialized tool focusing on integer overflow. EasyFlow uses taint analysis for overflow detection. Detection algorithm analyzes transaction instructions and mathematical instructions (at bytecode level) like EXP, ADDMOD, and MULMOD and even the instructions protected by SafeMath library. 

\item [--] {\it ContractFuzzer \cite{47}.} Fuzzing is a technique which can perform both static and dynamic analysis independently or at the same time. \cite{6} discusses an example of fuzzing with SC by mounting the Truffle project to a docker image. ContractFuzzer detects threats like ``Gasless send'' (section \ref{subsubsec:gaslessSend}), ``Exception disorder'' (section \ref{subsubsec:mshe_ucs}), Reentrancy (section \ref{subsubsec:reentrancy}), TSD (section \ref{subsubsec:tsd}), and Freezing Ether (section \ref{subsubsec:frozenEther}).

\end{itemize}

  \vspace{-0.1in}
\section{Conclusion and Research Directions}
\label{sec:conclusion}

We have provided classification of SC vulnerabilities but this can be further enhanced. In fact, security of SC is vital for strengthening the concept of BlockChain. Consistent efforts from academia have provided great solutions and this should continue. Our future research would be related to identifying randomization vulnerabilities. 
Following are the suggestion to fill the gaps in previous research and to advance the current research:

\begin{itemize}
    \item[--] For EVM researcher: a) there is a need for run-time environment within the Ethereum (i.e. EVM) and all the newly launched SC must be tested using this environment. This would take care of SC which are launched without testing, and b)	Many good tools have been developed by academia and it would be a good approach to incorporate them on the Remix website as plug-ins.

\item[--] For Solidity researcher: Solidity interpreter needs more improvements to catch the vulnerabilities. For catching mathematical errors, Solidity can incorporate a solver. This addition can also pave the way for detection of reentrancy error, which occurs due to the misplacement of account deduction statement

\item[--] For general researcher: Improving and developing new tools and techniques is important. Some examples areas where tools can be developed are: i) to catch new vulnerability patterns such as reported in \cite{83, 28}, ii) to make the use of libraries safe for SC, and iii) design strategies to reduce the miner's time to generate a block that is currently 900 seconds.  

\item[--] For Security Researchers: It is also important to develop more effective security testing and adequacy criteria that are unique for testing SC \cite{DBLP:conf/sac/DassN20}. It is also important to develop algorithmic \cite{DBLP:conf/compsac/DassN20} and machine learning \cite{DBLP:conf/bigdataconf/AbriSKSN19}, and deep learning \cite{DBLP:conf/compsac/Tavakoli20} approaches for detecting vulnerabilities and security defects in SCs.

\end{itemize}

\section*{Acknowledgment}
This research work is supported by National Science Foundation (NSF) under Grant No: 1821560.

\bibliographystyle{IEEEtran}
\bibliography{sourcefile-bib}

\end{document}

%% file: solidity-highlighting.tex
\usepackage{listings, xcolor}

\definecolor{verylightgray}{rgb}{.97,.97,.97}

\lstdefinelanguage{Solidity}{
	keywords=[1]{anonymous, assembly, assert, balance, break, call, callcode, case, catch, class, constant, continue, constructor, contract, debugger, default, delegatecall, delete, do, else, emit, event, experimental, export, external, false, finally, for, function, gas, if, implements, import, in, indexed, instanceof, interface, internal, is, length, library, log0, log1, log2, log3, log4, memory, modifier, new, payable, pragma, private, protected, public, pure, push, require, return, returns, revert, selfdestruct, send, solidity, storage, struct, suicide, super, switch, then, this, throw, transfer, true, try, typeof, using, value, view, while, with, addmod, ecrecover, keccak256, mulmod, ripemd160, sha256, sha3}, 
	keywordstyle=[1]\color{blue}\bfseries,
	keywords=[2]{address, bool, byte, bytes, bytes1, bytes2, bytes3, bytes4, bytes5, bytes6, bytes7, bytes8, bytes9, bytes10, bytes11, bytes12, bytes13, bytes14, bytes15, bytes16, bytes17, bytes18, bytes19, bytes20, bytes21, bytes22, bytes23, bytes24, bytes25, bytes26, bytes27, bytes28, bytes29, bytes30, bytes31, bytes32, enum, int, int8, int16, int24, int32, int40, int48, int56, int64, int72, int80, int88, int96, int104, int112, int120, int128, int136, int144, int152, int160, int168, int176, int184, int192, int200, int208, int216, int224, int232, int240, int248, int256, mapping, string, uint, uint8, uint16, uint24, uint32, uint40, uint48, uint56, uint64, uint72, uint80, uint88, uint96, uint104, uint112, uint120, uint128, uint136, uint144, uint152, uint160, uint168, uint176, uint184, uint192, uint200, uint208, uint216, uint224, uint232, uint240, uint248, uint256, var, void, ether, finney, szabo, wei, days, hours, minutes, seconds, weeks, years},	
	keywordstyle=[2]\color{teal}\bfseries,
	keywords=[3]{block, blockhash, coinbase, difficulty, gaslimit, number, timestamp, msg, data, gas, sender, sig, value, now, tx, gasprice, origin},	
	keywordstyle=[3]\color{violet}\bfseries,
	identifierstyle=\color{black},
	sensitive=false,
	comment=[l]{//},
	morecomment=[s]{/*}{*/},
	commentstyle=\color{gray}\ttfamily,
	stringstyle=\color{red}\ttfamily,
	morestring=[b]',
	morestring=[b]"
}